\documentclass[preprint,preprintnumbers,showkeys, amsmath,amssymb]{revtex4}
\usepackage{graphicx}
\usepackage{dcolumn}
\usepackage{bm}
\usepackage{epsfig}

\def\n{{\bf{\hat{n}}}}
\newcommand{\be}{\begin{equation}}
\newcommand{\e}{\end{equation}}
\newcommand{\bear}{\begin{eqnarray}}
\newcommand{\ear}{\end{eqnarray}}

\newcommand{\de}{{\rm d}}

\newcommand{\etar}{\eta_{\rm LSS}}
\newcommand{\valpha}{\boldsymbol{\alpha}}

\newcommand{\begm}{\begin{pmatrix}}
\newcommand{\enm}{\end{pmatrix}}
\def\apjs{Astrophys.J.Suppl}
\def\apj{ApJ}
\def\apl{ApJL}
\def\mnras{MNRAS}
\def\prd{PRD}
 
\def\mnras{MNRAS}
\def\HI{{\rm HI}}
\begin{document}

\title{CMBR Weak Lensing and HI 21-cm Cross-correlation Angular Power Spectrum} 
\author{Tapomoy Guha Sarkar}\email{tapomoy@cts.iitkgp.ernet.in} 

\affiliation{ Centre for  Theoretical Studies,  
Indian Institute Of Technology. Kharagpur, 721302, India}

\begin{abstract}
Weak gravitational lensing of the CMBR manifests as a secondary
anisotropy in the temperature maps. The effect, quantified through the
shear and convergence fields imprint the underlying large scale
structure (LSS), geometry and evolution history of the Universe. It is hence
perceived to be an important observational probe of
cosmology. De-lensing the CMBR temperature maps is also crucial for
detecting the gravitational wave generated B-modes. Future observations of redshifted 21-cm radiation from the cosmological neutral hydrogen (HI) distribution hold the potential of probing the LSS over a large redshift range. 
We have investigated the correlation between post-reionization HI
signal and weak lensing convergence field. Assuming that the HI follows
the dark matter distribution, the cross-correlation angular power
spectrum at a multipole $\ell$ is found to be proportional to the cold
dark matter power spectrum evaluated at $\ell/r$, where $r$ denotes
the comoving distance to the redshift where the HI is located. The
amplitude of the cross-correlation depends on quantities specific to
the HI distribution, growth of perturbations and also the underlying
cosmological model. In an ideal situation, we found that a statistically
significant detection of the cross-correlation signal is possible. If detected, the cross-correlation
signal holds the possibility of a joint estimation of cosmological
parameters and also may be used to  test various CMBR de-lensing estimators.
\end{abstract}

\maketitle

\section{Introduction}

Weak gravitational lensing \cite{ludoV}  of distant background sources by
intervening large scale structure, distorts their images over large
angular scales. The effect arises due to fluctuations of the
gravitational potential, and a consequent deflection of light by
gravity.  Measurement and quantitative study of these distortions
allows  us  to probe the matter distribution and geometry of the universe. 
Late time evolution of the universe is dictated by dark energy through
a modification of the growing mode of perturbations or through
possible clustering properties of dark energy ( $ w \neq -1$ ).
Weak lensing studies can be used  to  impose constraints on various
cosmological parameters and hence, implicitly probe dark energy models
\cite{hoekjain} and modified gravity theories \cite{mog}. It is
relevant for our present purpose to note that weak lensing is directly
related to the underlying matter distribution of the universe.
Weak lensing of background source galaxies by large scale structure
(cosmic shear) has been studied extensively, and the measurements have 
been used for projected mass reconstruction (for review \cite{munshi}) .

Gravitational lensing also deflects the photons which are free streaming from
the last scattering surface (epoch of recombination $ z \sim 1000$)
and  manifests as a secondary anisotropy in the Cosmic microwave background radiation (henceforth CMBR) brightness temperature maps \cite{antony}.
Despite, the intrinsic weakness of the `signal to noise ratio' for the above
effect, weak lensing of CMBR can, in principle be used to probe the
universe at distances ($z \sim 1100$) much larger than any galaxy-
redshift surveys. Moreover CMBR lensing studies do not face the
problems  arising due to intrinsic alignment of
source galaxies.
Standard techniques to measure  secondary anisotropies in CMBR,
 uses the cross correlation of some relevant
observable (related to the CMB fluctuations) with fluctuations of some
tracer of the large scale structure \cite{hirata,smith, padma}.
Observables relevant to weak lensing are `Convergence' and the `Shear'
fields,  which quantify the distortion of an image due to
gravitational lensing.
Convergence ($\kappa$)
measures the lensing effect through its direct dependence on
the gravitational potential and it probes geometry implicitly through its
dependence on  various cosmological distances.
 
Future experiments (PLANK
\footnote{http://www.rssd.esa.int/index.php?project=planck}, CMBPOL \cite{cmbpol} etc) would provide high resolution
maps for the CMB temperature and polarization fields. The effect of
gravitational lensing can be extracted from these maps by constructing 
various estimators for the convergence field ($\kappa$) through
quadratic  combination of these fields (T, E, B) \cite{huapj, seljak, huoka}. One could also predict
the noise involved in such estimation based upon various experimental 
parameters. Lensing reconstruction can also be done using the $21$ cm
observations \cite{zahnzalda}. The reconstructed convergence field can then be used for
cross correlation.
De-lensing the CMB maps is also crucially important for detecting the
gravitational wave generated B-mode.

It is well accepted that the  the neutral hydrogen (henceforth HI)
 distribution in the post-reionization epoch ($ z \lesssim 6$) largely
 traces  the underlying large scale structure of the universe
 \cite{bali, furla, lewcha}. This  allows us to relate HI distribution
 to the cold dark matter  distribution through a  possible
 `bias'. Matter perturbations are in  the  linear regime on large
 scales under consideration and  the above simplifying
 assumption is reasonable.
 Hence, observations of the redshifted $21 \,{\rm cm}$ radiation of the HI
spin-flip hyperfine transition provides an unique opportunity for
 probing the universe over a wide range of redshifts ($200 \ge z \ge
 0$) \cite{bali,furla, lewcha}. 
Theoretical predictions \cite{bns,bs} have suggested the use of HI,
statistically, as a probe of large scale structure. 
Positive correlation between the optical
 galaxies (6dFGS) and  HI fluctuations \cite{Pen et al} has also been
 observed recently. 
        
In this paper we have investigated the possibility of using  diffused
cosmological HI as a tracer of the underlying large scale structure to
probe weak lensing induced secondary anisotropy of the
CMBR. 
Cosmic shear fields imprint the underlying distribution of matter
over large scales. 
We have studied the the cross correlation between the
post-reionization fluctuations  in the HI brightness temperature and the
weak lensing convergence field. The cross-correlation angular power spectrum,  measures the strength of the
correlation as a function of the angular scale.

The weak lensing of CMBR, quantified through the convergence field is
expressed as a line of sight integral.
 Cross correlation of weak lensing with  the HI fluctuations, however pick up the 
 contribution from only one redshift ($z_{HI}$ at which the HI is
 probed). 
The advantage of using HI observations is that, the redshifted   $21
\, {\rm cm}$ line emission observations  allow us to probe the
universe continuously at different redshifts.
  We can probe the 
integral effect of weak lensing at any intermediate redshift by suitably tuning
the frequency band for HI observation. This, in principle enables us
to do a tomographic study of the late-time cosmic history continuously
over an entire range of redshifts.
On similar lines, cross-correlation of HI temperature map with the CMBR, aimed to
isolate the ISW signal (an integral effect) has been studied \cite{tapomoy}.

Several Radio telescopes (eg.currently functioning GMRT
\footnote{http://www.gmrt.ncra.tifr.res.in/} and upcoming  MWA 
\footnote{http://www.haystack.mit.edu/ast/arrays/mwa/} \& LOFAR
\footnote{http://www.lofar.org/}) 
 are aimed to map the cosmological distribution of HI  at high redshifts. 
The extreme weakness of the post-reionization HI signal ($ < 10
\,\mu{\rm Jy}$) from individual clouds, despite some magnification due to  Gravitational
lensing  \cite{saini},  poses a serious observational challenge.
However, observation of the statistical distribution of HI as  a weak 
background in radio observations does not require
 the need to resolve individual galaxies. Such observations contain  information
about the HI fluctuations at the comoving distance being probed (frequency) \cite{bns,bs}.

Convergence field reconstructed from CMBR maps of large portion of the
sky  and a corresponding HI map would allow us to compute the
cross-correlation power spectrum and hence independently quantify the
cosmic history at redshifts $z \le 6$. The cross-correlation power
spectrum  may also independently  compare the various theoretical estimators that
separate the lensing contribution from the CMB data .

\section{Formulation}

The lensed CMB brightness temperature $\tilde{T}({\n})$ along the  direction of
the unit vector $ \n$ is related to the unlensed
temperature $T({\n})$ through the map $ \tilde{T}({\n}) = T(\n +{
  \valpha})$,
where $\valpha $ denotes the total deflection due to weak lensing by
the intervening large scale structure.
At the lowest order, magnification of the signal is given by the
convergence,  $\kappa = -\frac{1}{2}\nabla\cdot\valpha$.The convergence field can be written as a line of sight integral given
by \cite{ludoV}
\be
\kappa(\n) = \frac{3}{2} \Omega_{m0} {\left( \frac{H_0}{c} \right)}^2
\int_{\eta_0}^{\eta_{LSS}} d\eta F(\eta) \delta(d_{A}(\eta)\n , \eta)
\e
where $d_{A}$ stands for the comoving angular diameter distance and $F(\eta)$ is
given by 
\be
F(\eta) = \frac {d_A(\eta_{LSS} - \eta)d_A(\eta)D_{+}(\eta)}{d_A(\eta_{LSS}) a(\eta)}
\e
Here $D_{+}$ denotes the growing mode for the density contrast
$\delta$, and $\eta_{LSS}$ denotes the conformal time corresponding to the
last scattering surface (assuming instantaneous recombination), and
$a(\eta)$ denotes the scale factor.

Here we have excluded weaker contribution to the convergence field from
sources other than large scale structure (like gravitational waves).
Expanding this in  the basis of spherical
harmonics 
\be
\kappa(\n) = \sum_{\ell,m}^\infty a_{\ell m}^{\kappa} Y_{\ell m}({{\n}})  
\e

The expansion coefficients $a_{\ell m}^{\kappa}$  can be obtained by
integrating over the solid angle $\omega_{\n}$  as
\be
a_{\ell m}^{\kappa} = \int d\omega_{\n} \kappa(\n) Y_{\ell m}^*({\n}) 
\e

Using the Raleigh expansion
\be
 e^{i{\mathbf{k}}\cdot {\mathbf{n}} r}  = 4\pi \sum_{\ell,m}{
   (-i)}^{\ell} j_{\ell} (kr)Y_{\ell m}^*({\bf{\hat{k}}}) Y_{\ell m}({\bf{\hat{n}}})
\e
we have
\be
a_{\ell m}^{\kappa}
= 6\pi\Omega_{m0} {\left( \frac{H_0}{c} \right)}^2 {(-i)}^{\ell} \int
\frac{d^3{\mathbf{k}}}{{(2\pi)}^3} 
 \int_{0}^{\eta_0} \de \eta \,
F(\eta)\delta({\mathbf{k}}) j_{\ell}(kr)Y_{\ell m}^*({\bf{\hat{k}}}) 
\label{eq:a3}
\e

where $ \delta({\mathbf{k}})$ 
is the Fourier transform of $ \delta({\mathbf{r}})$, and $j_\ell(x)$
is the spherical Bessel function.

In studying the post-reionization HI power spectrum we assume  that the HI traces the underlying dark matter 
distribution with a possible bias function ${b}(k) = {[P^{HI}(k)/ P(k)]}^{1/2}$, where  ${P}^{HI}(k)$ and ${P}(k)$ denote the HI and
  dark matter power spectra respectively. This function is assumed to quantify the clustering
property of the neutral gas. It is  believed that, on small
scales (below the Jean's length), the linear density contrast for the
gas is related to the dark matter density contrast though a scale
dependent function \cite{fang}. However the bias is known to be  reasonably
scale-independent on large scales. The length scale above which the
bias is linear,  depends crucially on the redshift being probed.
Numerical simulations indicate that the large scale  linear bias grows
monotonically with redshift for $1< z< 4$ \cite {marin}. This is known to be
true for galaxies \cite{fry, mo, moo}.
The increase in the amplitude of HI brightness temperature power spectrum
is however slow (a factor of $\sim 2$  for z between $1$ and $5$)\cite{bagla}. 
In this paper we have considered scales which are much larger than the
scale of non-linearity and hence linear scale independent bias has been used.

Expanding the HI 21-cm brightness temperature fluctuations
(in Fourier space \cite{bharad04}) from redshift $z_{HI}$  in terms of spherical harmonics and proceeding
as before we get 
\be
a_{\ell m}^{\rm HI}   {=}  4\pi \bar{T}(z)\bar{x}_{HI} {(-i)}^{\ell}
\int
\frac{d^3{\mathbf{k}}}{{(2\pi)}^3} 
\delta({\mathbf{k}},a)J_{\ell}(kr)Y_{\ell m}^*({\bf{\hat{k}}}) \,.
\label{eq:a4}
\e

where  $\bar{x}_{HI}$  is the mean HI  fraction, and 
\be
\bar{T}(z)=4.0 \, {\rm mK}\,\,(1+z)^2  \, \left(\frac{\Omega_{b0}
  h^2}{0.02}\right)  \left(\frac{0.7}{h} \right) \frac{H_0}{H(z)}
\e
The term  $\mu ={\bf{\hat{k}}}\cdot{\bf{\hat{n}}}$ has its origin in the  HI peculiar
velocities \cite{bns,bharad04} which have also been assumed to be
caused by the dark matter fluctuations.
In equation (\ref{eq:a4}) we have defined
\be
J_{\ell}(x) {=}  b j_{\ell}(x)- f \frac{d^2 j_\ell}{dx^2} \,.
\e
Where $f$ denotes the logarithmic derivative of the growing mode and
is given by $ f = \Omega_m ^ {0.6}$.

At redshifts $0 \le z \le 3.5$  we have 
$\Omega_{\rm gas} \sim 10^{-3}$ (for details see
 \cite{peroux,lombardi,lanzetta}). This allows us to calculate the 
 mean neutral
fraction of the hydrogen gas $ \bar{x}_{\HI}=50\,\,\Omega_{\rm gas}
h^2 (0.02/\Omega_b h^2) =2.45 \times 10^{-2}$, which we assume is a
 constant over the entire redshift range $0 \le z \le 6$.

We use equations (\ref{eq:a3}) and  (\ref{eq:a4}) to calculate
$\mathcal{C}^{HI-\kappa}_{\ell}$,  the
cross correlation angular power spectrum between the 
HI 21-cm brightness temperature signal and the convergence field, defined
through 
\be
\langle  a_{\ell m}^{\kappa} a_{\ell'm'}^{*\rm HI} \rangle  =
\mathcal{C}^{HI-\kappa}_{\ell} 
\delta_{\ell \ell'}\delta_{mm'} 
\e
Note that $\mathcal{C}^{HI-\kappa}_{\ell}$ also depends on $z_{HI}$,  the
redshift from which the HI signal originates, or equivalently on
$\nu=1420 \, {\rm MHz}/(1+z_{HI})$,  the frequency of  the HI
observations (not explicitly mentioned here).   

We obtain 
\be
\mathcal{C}_{\ell}^{HI-\kappa} {=} {A}(z_{HI}) \int dk \left
	[k^2  P(k)
  J_\ell(k r_{HI}) \int_{\eta_0}^{\etar} \de \eta F(\eta)
  j_{\ell}(kr)\right]
\e 

where $P(k)$ is the present day dark matter power spectrum, 
\be
 {A}(z) = \frac{3}{\pi}\Omega_{m0} {\left( \frac{H_0}{c} \right)}^2 \bar{T}(z)\bar{x}_{HI}
 D_{+}(z)    
\e

For large $\ell$ (small angular scales where ``flat sky''
  approximation is reasonable) the Limber approximation in Fourier space \cite{limber,Afshordi}, $
j_{\ell}(kr) \approx  \sqrt{\frac{\pi}{2\ell+1}} \delta_D(\ell + \frac{1}{2} - kr)
$, allows  us to understand various generic  scaling properties of  the angular cross-correlation power spectrum. 
\be
   \mathcal{C}_{\ell}^{HI-\kappa}  \propto \frac{\pi}{2}  {A}(z_{HI})  \frac{F(z_{HI})}{{d_{A}(z_{HI})}^2} 
P{\left(\frac{\ell}{r_{HI}}\right)}
\label{eq:fnl}
\e
where $P(k)$ is the present day dark matter power spectrum and all the
 terms on the {\it rhs.} are evaluated at 
$z_{HI}$.

 Using equation (\ref{eq:a3}) we have the Convergence auto-correlation
power spectrum which for large $\ell$ can be approximately
written as
\be
\mathcal{C}_{\ell}^{\kappa} \approx \frac{9}{4}\Omega_{m0}^{2} {\left(
  \frac{H_0}{c} \right)}^4 \int d \eta\frac{ F^2(\eta) }{d_A^2(\eta)}
P{\left(\frac{\ell}{d_A(\eta)}\right)}
\e

We also have, for comparison, the  HI-HI angular power spectrum
$\mathcal{C}_{\ell} ^{HI}(z_{HI}) $ 
\cite{datta1}, which  describes the statistical properties of HI
fluctuations . 

The function $\mathcal{C}_{\ell}^{HI}(z_{HI})$ is known to be a  direct observational
 estimator of the HI fluctuations at redshift $z_{\HI}$ and  does not
 require the assumption of an underlying cosmological model (eg. \cite{ali08}).
Using the `flat sky' approximation 
\cite{datta1}, which is reasonable for $ \ell > 10 $, we have
$ \mathcal{C}_l^{HI} (z_{HI}) $  given by\
\begin{equation}
\mathcal{C}_\ell^{HI}(z_{HI} ) =
\frac{\bar{T}^2~ }{\pi r_{\nu}^2} \bar{x}^2_{\HI}   D_{+}^2
\int_{0}^{\infty} {\rm d} k_{\parallel} \, 
 \left[  b+  f \mu^2 \right]^2 P(k) \,
\label{eq:fsa} 
\end{equation}
where $ r$ is the comoving distance corresponding to the
redshift $z_{HI}$ or equivalently frequency $\nu = 1420 {\rm
  MHz}/(1+z_{\HI})$,  and $k=\sqrt{k^2_{\parallel}+(l/r)^2}$. 
In this paper we have used the WMAP$5$ data for the various
cosmological parameters.

 We note  that the quantity of interest -  the convergence field
 $\kappa(\n)$,   is
not a direct observable in CMBR experiments. The degree of non-gaussianity in the lensed CMB maps is proportional to the lensing
potential responsible for it. This allows a reconstruction of the weak
lensing potential and consequently the deflection angle
$\vec{\alpha}$, through the
use of various statistical estimators \cite{hanson, kesden}. The reconstructed lensing
 convergence field is sensitive to the statistical tool (estimator)
 being used and reflects the degree of de-lensing achieved.

 The estimated quantity, we have focussed on,  namely the cross correlation angular  power spectrum, $ \mathcal{C}_{\ell}^{HI-\kappa}$, does not directly de-lens the CMB
 maps. It however  uses the reconstructed convergence field, and is hence sensitive to the underlying  de-lensing
 technique, and the cosmological model.
We have calculated the theoretical cross-correlation power spectrum
 assuming a standard cosmological model. The estimated  $\mathcal{C}_{\ell}^{HI-\hat{\kappa}} $, (where $\hat{\kappa}$ is the
 estimated convergence field)  with its known error bars can be
 compared with our  predicted $ \mathcal{C}_{\ell}^{HI-\kappa}$. 
Hence, the theoretical cross-correlation angular power spectrum  provides a
 template to independently compare various  estimators which are  aimed at de-lensing the CMB maps.

\section{Results}

\begin{figure}[h]
\begin{center}
  \mbox{\epsfig{file=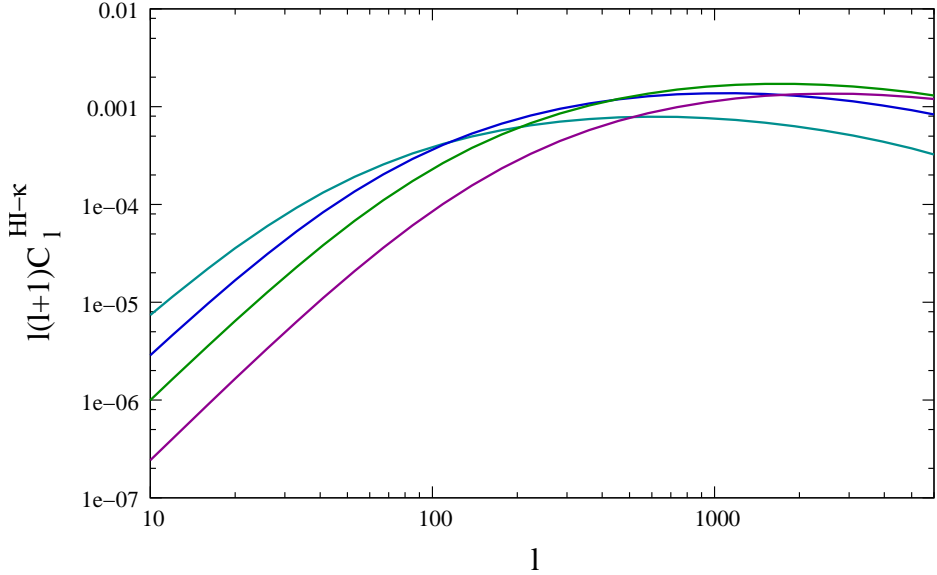,width=0.75\textwidth,angle=0}}
\caption{The HI- convergence angular power spectrum  for  redshifts $z=
  0.5. 1.0, 2.0$ and $ 5.0$ (top to bottom).} 
\label{fig:kappa}
\end{center}
\end{figure}

\begin{figure}[h]
\begin{center}
  \mbox{\epsfig{file=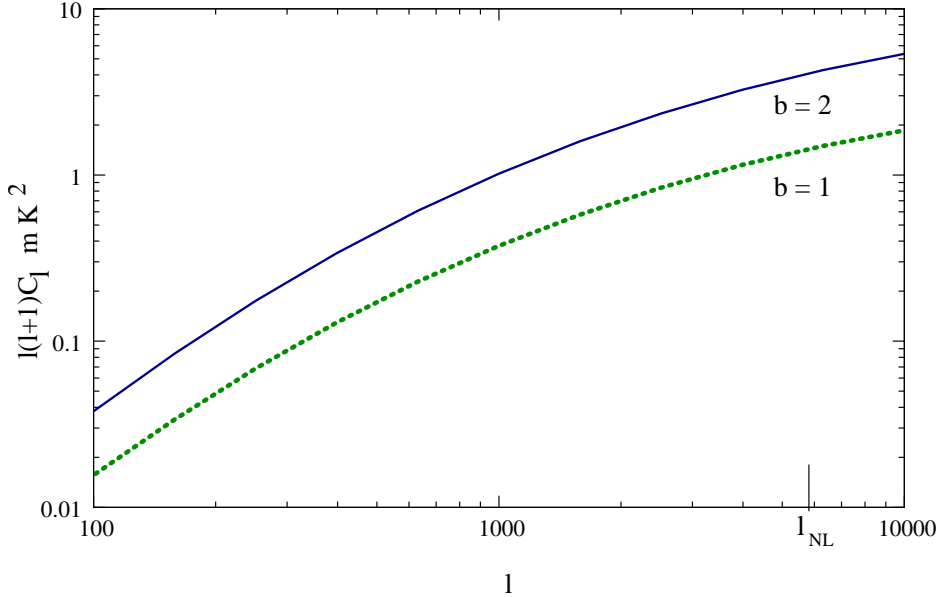,width=0.75\textwidth,angle=0}}
\caption{The HI angular power spectrum  at redshift $ z=
  3.3$  showing the effect of linear bias. $ l_{\rm{NL}}$ is the scale
  above which non-linear biasing should be incorporated }
\label{fig:bias}
\end{center}
\end{figure}

\begin{figure}[h]
\begin{center}
\mbox{\epsfig{file=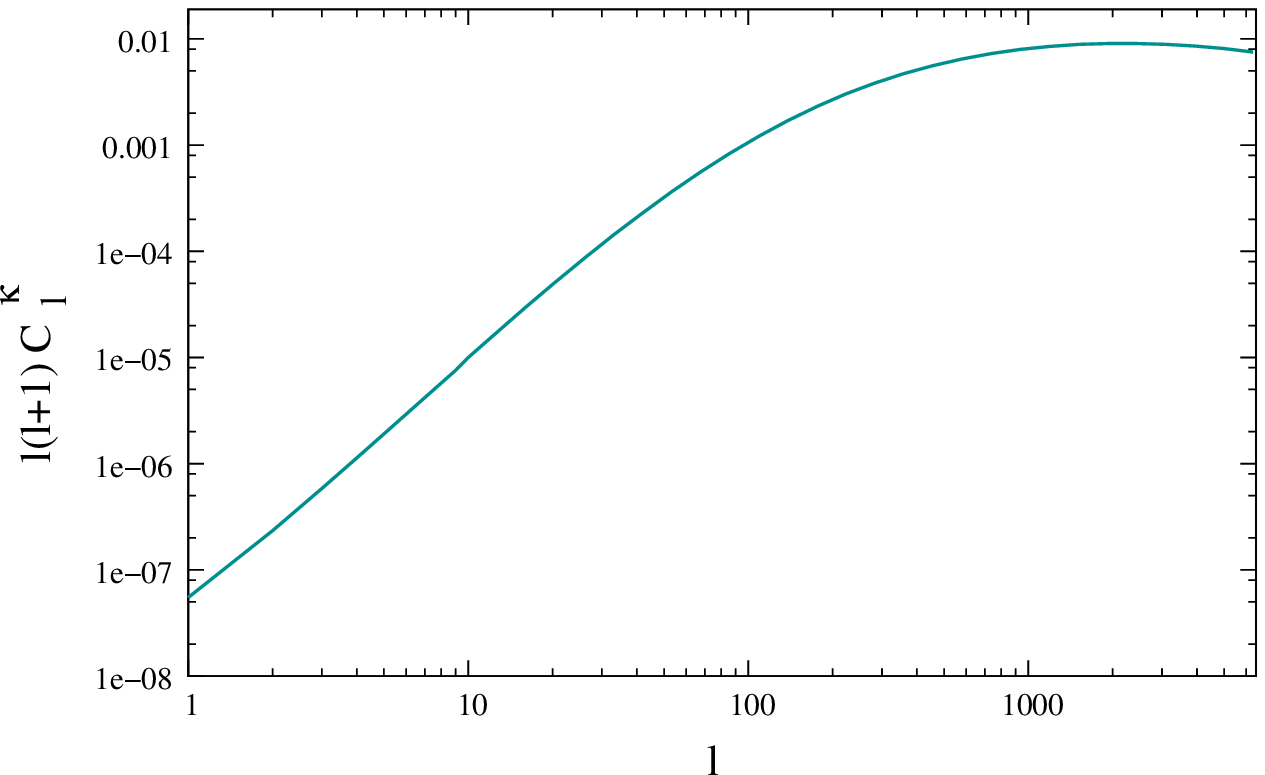,angle=0}}
\caption{The convergence power spectrum  $C_\ell^{\kappa}$ }
\label{fig:conv}
\end{center}
\end{figure}

Figure \ref{fig:kappa} shows the theoretically predicted
cross-correlation angular power spectrum
$\mathcal{C}_{\ell}^{HI-\kappa} $ for various redshifts $0.5 \le
z_{HI} \le  6$. The currently favored  $\Lambda$CDM
 cosmological model 
 with parameters  $(\Omega_{m0},\Omega_{\Lambda0},h,\sigma_8,n_s) = $   ($ 0.28$, $0.72$, $0.7$,
 $0.82$, $0.97$) \cite{tegmark,komatsu} has been used here. For HI
 signal we have assumed a linear bias model (reasonable on the large
 scales under consideration) with $ b = 1$ in the fiducial model.

Numerical simulations have indicated the deviation from $ b = 1 $ at 
high redshifts. It is seen that at large scales the linear bias is  
$b \sim 2$ for $ z \sim 3$. The effect of larger (scale independent) bias is shown in figure
\ref{fig:bias}.
Apart from the scaling of the power spectrum at large scales the bias
also has a weak effect of modifying the power spectrum amplitude
through the change in the  the redshift space
distortion factor $ \beta = f/b  $. We have also indicated the scale  $l_{NL} \sim k_{\rm{NL}}
r_{z}$,  above which the linear bias assumtion is invalid. For  $z \sim
3 $ this angular scale $ \l_{\rm{NL}} \sim 6000$. We have restricted ourselves
to multipoles less than that.
 
Figure \ref{fig:conv} shows the Convergence auto-correlation power
spectrum for reference.
The Cross-correlation power spectrum has the same shape as the matter
power spectrum.  For different redshifts the signal peaks at a
particular $\ell$ which  scales as $ \ell \propto  r_{HI}$. The
angular distribution of power clearly follows the underlying
clustering properties of matter.
The amplitude of the cross-correlation power spectrum depends on
various factors some of which are related to the underlying
cosmological model and others related to the HI distribution at $ z_{HI}$.
The angular diameter distances directly imprint the geometry of the universe 
and also depends on the cosmological parameters.
The $21cm$  signal has been proposed to be an useful
probe of the cosmological parameters \cite {BSS, wyithe,mcquin}. The
cross-correlation signal may likewise be used independently for joint estimation of parameters.

We shall now discuss the prospect of detecting the cross-correlation
signal. 
Redshifted $21$ cm signal is buried deep under foregrounds. Removal of
the foreground component is a major challenge
\cite{ali08,datta1,mcquin,santos}. However, it is to be noted that cross-correlation between the HI brightness temperature field and
the convergence field is much less likely to be affected by
foregrounds or other systematics.
The error in the cross-correlation signal is a sum in quadrature, of
the contribution due to instrumental noise and sample variance.
Increased resolution (for CMB experiment) and increased time of
observation (for 21 cm observation) can in principle significantly
reduce the instrumental noise. Sample variance however puts a
fundamental bound on the detectability of the signal. 

The sample variance for the cross-correlation angular power spectrum  
$\mathcal{C}_{\ell}^{HI-\kappa} $ is given by
\be
\sigma_{SV}^2 = \frac {\mathcal{C}_{\ell}^{\kappa}
 \mathcal{C}_{\ell}^{HI}}{(2\ell + 1){\sqrt{N_c}}f_s \Delta\ell}
\e
Where the numerator contains the auto-correlation angular power spectra. $\Delta\ell$
represents a band in $\ell$ and $ f_s$ is fraction of sky common to
the convergence field $\kappa$ and HI brightness temperature
distribution $\frac{\Delta T}{T}$. $N_c$ denotes the number of
independent estimates of the $21$cm observations obtained from
different frequency channels in a given frequency band and suppresses
the sample variance by a factor $ 1/\sqrt{N_c}$.

\begin{figure}
\begin{center}
\mbox{\epsfig{file=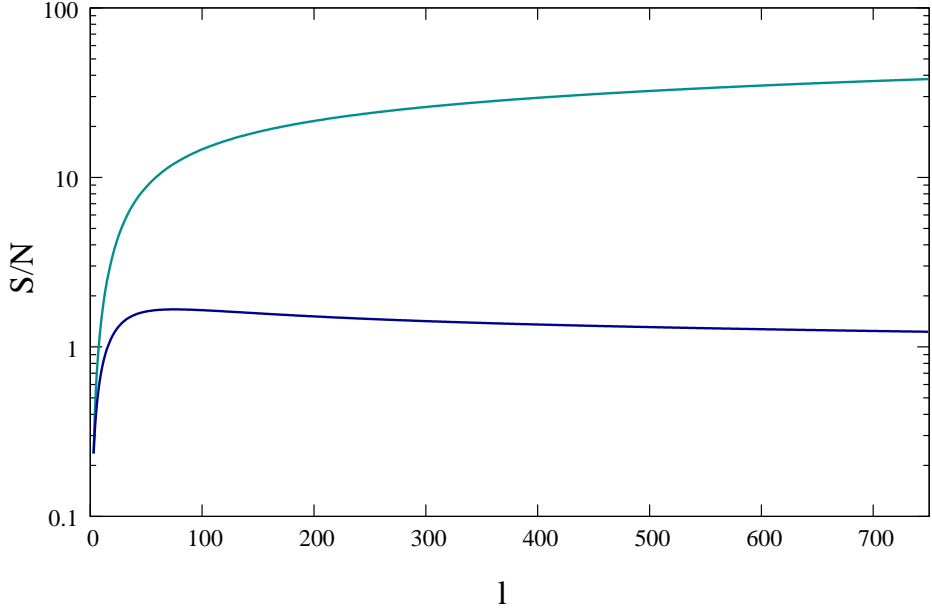,width=0.75\textwidth,angle=0}}
\caption{ The lower curve shows Signal to Noise ratio (S/N) as a function of
  angular scale $\ell$. The upper curve shows the effect of
  summing over multipoles. The probing redshift $z_{HI} = 0.5$ }
\label{fig:sn}
\end{center}
\end{figure}

We have used the ideal hypothetical  possibility of $f_s = 1$, and
used $\Delta\ell = 1$ . we have chosen $N_c = 32 $  assuming that the HI signal decorrelates
over a frequency separation of $\sim 1 \rm MHz $ and hence yield $32$
independent estimates  for  a  $ 32    \rm   MHz$ bandwidth radio
observation.
The estimated  Signal to Noise ratio $ S/N
=\mathcal{C}_{\ell}^{HI-\kappa}/\sigma_{SV} $ is shown in fig \ref{fig:sn} . for
$z_{HI} = 0.5 $. The predicted $ S/N$ is seen to be $\sim 2$ and is not high
enough for a statistically significant detection which requires $S/N
\geq 3$. Choosing a  $\Delta\ell = 10$ for $\ell \leq 100$ and
$\Delta\ell = 100$ for $ \ell > 100$ will however produce a $S/N > 3$.

It is possible to increase the $S/N$ by collapsing the signal from
different scales $\ell$ and thereby test the feasibility of a
statistically significant detection. The Signal to Noise cumulated upto a multipole
$\ell$ is defined as  (see \cite{cooray2} for similar calculation)

\be
{\left(\frac{S}{N} \right)}^2 = \sum \frac{(2\ell + 1) \sqrt{N_c}
  f_s{(\mathcal{C}_{\ell}^{HI-\kappa})}^2
}{(\mathcal{C}_{\ell}^{HI} + N_{\ell}^{HI}
  )(\mathcal{C}_{\ell}^{\kappa} + N_{\ell}^{\kappa}) }
\e
The summation in the above equation extends up to a certain $\ell$.
$ N_{\ell}^{\kappa}$ and $ N_{\ell}^{HI}$ denotes the noise power
spectrum for $\kappa$ and HI observations respectively.
Ignoring the instrument noises we note  that there is a significant increase in the $S/N$ by cumulating over multipoles $\ell$.
This implies that a statistically significant detection of
$\mathcal{C}_{\ell}^{HI-\kappa} $ is possible and the signal is not
cosmic variance limited.
$21$- cm observations allow us to probe a continuous range of
redshifts. This allows us to further increase the $S/N$ by collapsing
the signal from various redshifts.
As discussed earlier, an increased HI bias would   increase the
signal. However the $ S/N \propto C^{HI-\kappa} / \sqrt {C^{HI}} $  is
not expected to be  seriously affected.  

Instrumental noise plays an important role at large multipoles (small
scale).
For a typical CMB experiment,  the noise power spectrum  \cite{mar,huu} is given by
$
N_{\ell} = \sigma^2_{\rm{pix}}\Omega_{\rm{pix}}{ W_{\ell}}^{-2}
$, where different pixels are assumed to have uncorrelated noise with
uniform variance $\sigma^2_{\rm{pix}}= s^2/ t _{\rm{pix}} $, where $ s^2$
and $t_{\rm{pix}}$  denotes pixel sensitivity and `time spent on the
pixel' respectively. $\Omega_{\rm{pix}} $ is the solid angle subtended
per pixel and we choose a gaussian beam $ W_{\ell} = \rm{exp} [ { - \ell^2
  \theta^2_{FWHM}/ 16  ln 2}] $.
For CMBPOL \cite{cmbpol} like experiments, the noise power spectrum
for $\kappa$ with the beam  FWHM $ \sim  3'$   and sensitivity $\sim 1 \mu K -
\rm{arcmin} $
 is  $  N_{\ell}^{\kappa} < 10^{-8}$ for $ \ell < 3000 $  (see \cite{sudeep,cmbpol})
Hence, $ N_{\ell}^{\kappa} \ll  \mathcal{C}_{\ell}^{\kappa}$ and maybe
ignored in our present analysis.

For HI observations, the quantity of interest is the complex Visibility
which is used to estimate the power spectrum \cite{ali08}.
For a radio telescope   with N antennae,  system temperature  $
T_{sys}$,  operating at a  frequency
$\nu$, and  band width $B$ the  noise correlation is given by \cite{baglakanan}
\be
N_{\ell}^{HI} = \frac{4}{\sqrt{2 \pi}N ( N - 1)}{ \left[
    \frac{T_{sys}}{K} \right ]}^2 \frac{1}{T \sqrt{{\Delta \nu}
    B}}\frac{1}{U^{0.5} \Delta U ^{1.5} \rho ( U ,\nu)}
\e

Where $ 2 \pi U \sim \ell $, $T$ denotes total observation time, and
    $K$ is related to the effective collecting area of the antenna
    dish . The function $\rho(U, \nu)$ takes  any non-uniform
    distribution of baselines into account and depends on
    the array design.
The bin $\Delta U  = 1/ \pi \theta_0$ is chosen assuming a gaussian
    beam of width $\theta_0$.
With a GMRT or MWA like instrument \cite{ali08},
one can in principle achieve a noise level much lesser than the
signal by increasing the time of observation (a 2000 hour observation
    is sufficient even with the present GMRT cofiguration)  and also by increasing
the band width of the instrument. Being  inversely related
to the number of antennae in the array,  future designs can allow
further suppression of the the system noise and achieve $ N_{\ell}^{HI} << \mathcal{C}_{\ell}^{HI}$.

This  establishes  the detectability of the cross-correlation signal.
We would like to conclude by noting that this theoretical prediction 
of positive correlation between weak lensing fields and 21 cm maps, 
 quantified through $  \mathcal{C}_{\ell}^{HI-\kappa}$ may allow an independent means to estimate various
 cosmological parameters and also test various estimators for CMBR delensing. 
 
\section{Acknowledgments}

T.G.S would like to acknowledge Somnath Bharadwaj for useful
discussions and help. Authors also acknowledge financial support from
the Board of Research in Nuclear Sciences (BRNS), Department of Atomic
Energy (DAE), Government of India.

\bibliography{apssamp}
 
\newpage


\end{document}